\documentclass{emulateapj}

\usepackage{graphicx}
\usepackage{epstopdf}
\usepackage{epsfig}
\shorttitle{A Double-Peaked Emitter in Binary Black Holes}
\shortauthors{Tang \& Grindlay}

\begin{document}


\title{The Quasar SDSS J153636.22+044127.0: A Double-Peaked Emitter in a Candidate Binary Black-Hole System} 


\author{Sumin Tang and Jonathan Grindlay}
\affil{Harvard-Smithsonian Center for Astrophysics, 60 Garden St,
Cambridge, MA 02138}


\begin{abstract}
Double-peaked emission lines are believed to be originated from
accretion disks around supermassive black holes (SMBHs), and about
3\% of $z<0.33$ AGNs are found to be double-peaked emitters. The
quasar SDSS J153636.22+044127.0 has recently been identified with
peculiar broad-line emission systems exhibiting multiple redshifts.
We decompose the H$\alpha$ and H$\beta$ profiles into a circular
Keplerian disk line component and other Gaussian components. We
propose that the system is both a double-peaked emitter and a binary
SMBH system, where the extra-flux in the blue peaks of the broad
lines comes from the region around the secondary black hole. We
suggest that such black hole binary systems might also exist in many
known double-peaked emitters, where the tidal torques from the
secondary black hole clear the outer region of the disk around the
primary black hole, similar to the gap in a protostellar disk due to
the process of planetary migration, and might also stimulate the
formation of a vertical extended source in the inner region around
the primary which illuminates the disk. However, most secondary
SMBHs in such systems might be too small to maintain a detectable
broad line region (BLR), so that the disk line from the primary
dominates.

\end {abstract}


\section{Introduction}
Boroson \& Lauer (2009) recently searched the Sloan Digital Sky
Survey (SDSS) archived quasar spectra (Schneider et al. 2007), and
identified a quasar, SDSS J153636.22+044127.0 (hereafter SDSS
J1536+0441), with two broad-line emission systems separated in
velocity by 3500 km/s. They interpreted it as a sub-parsec
supermassive black hole (SMBH) binary, where the two components in
the broad emission lines are from the two Broad Line Regions (BLRs)
around the pair. The existence of close SMBH binaries are expected
as the result of mergers of galaxies, and the sub-parsec scale
separation is significant, where the two black holes will eventually
coalescence via gravitational radiation (Begelman et al. 1980).

Subsequently, several observations have been carried out to study
the system, which also led to other interpretations (Chornock et al.
2009; Wrobel \& Laor 2009; Lauer \& Boroson 2009; Decarli et al.
2009). The most attractive interpretation is as an unusual
double-peaked emitter where the multiple components in the broad
lines are from the accretion disk around the black hole, which is
supported by a third velocity component (a ``bump'') of the broad
emission lines in the red wing (Chornock et al. 2009; Gaskell 2009).
Double-peaked emission lines in AGNs have profiles much like the
double-peaked emission lines of cataclysmic variables which are due
to the disk (Young \& Schneider 1980). However, only about 3\% of
$z<0.33$ AGNs are found to be double-peaked emitters in SDSS quasar
sample (Strateva et al. 2003), and about 20\% AGNs are found to be
double-peaked emitters in a radio-loud AGN survey which includes 106
AGNs (Eracleous \& Halpern 2003). By comparing the spectra taken at
different epochs, both Chornock et al. (2009) and Lauer \& Boroson
(2009) found no significant velocity evolution of the two main peaks
over $\sim 0.7$ years in the rest frame of the object, with a
$2-3\sigma$ upper limit of about $75-80$ km/s. Lauer \& Boroson
(2009) showed that the upper limit is still consistent with binary
black hole models, and none of the periods longer than $\sim200$
years has been ruled out.

Both models have their merits, but cannot explain the observed broad
line profile well alone. The two shoulders of double-peaked line
profiles, which are currently believed to be more likely to come
from accretion disks around single black holes (Eracleous et al.
1997), were first suggested to come from two BLRs associated with
the black holes of a supermassive binary (Gaskell 1983). However,
SDSS J1536+0441 has three velocity components in the H$\alpha$ and
H$\beta$ profiles, i.e. two sharp peaks and a shoulder-like bump in
the red wing, and therefore cannot be explained in the SMBH binary
model in Boroson \& Lauer (2009) which only produces two velocity
components. On the other hand, the strong sharp blue peaks have no
obvious analogue with other double-peaked emitters and thus require
a source other than a normal accretion disk.

However, the two explanations may fit together well. Accretion onto
AGN will usually be through an accretion disk, while SMBH binaries
are expected as normal consequences of galaxy mergers, therefore
combining the two together is reasonable. Here we propose an
alternative hypothesis that the system is both a double-peaked
emitter and a binary black hole system, where the extra-flux in the
blue peaks of the broad lines comes from the region around the
secondary black hole, and the disk lines are from the accretion disk
around the primary black hole. In this paper, we refer to the Broad
Line Region, or BLR, as any region producing the broad lines, which
may then include both the disk and discrete clouds. In section 2, we
fit the H$\alpha$ and H$\beta$ line profiles using a circular
Keplerian disk model. In section 3, we describe our model to explain
the system. We then discuss in section 4 that SMBH binary systems
might also exist in many other double-peaked emitters.

\section{Disk model fit}
In about half of the known double-peaked emitters, the broad
double-peaked lines could be fit well by a circular, symmetric,
relativistic, Keplerian disk ($60\%$ in Eracleous \& Halpern 2003
and $40\%$ in Strateva et al. 2003), while others require some form
of asymmetry (hot spots, elliptical disks, warps, spiral shocks,
etc.). Since we do not have any prior knowledge that the disk is
asymmetric and if so how it would be, we chose the axisymmetric
Keplerian disk model from Chen \& Halpern (1989), which has five
parameters: the inclination angle $i$, inner and outer radii $r_1$
and $r_2$ in units of the gravitational radius $r_G$, local
broadening of the line which is represented by a Gaussian rest-frame
profile of velocity dispersion $\sigma$, and the index of the
surface emissivity power law $q$. We fixed $q=-3$, as predicted in
photonionization calculations (Collin-Souffrin  \& Dumont 1989;
Dumont \& Collin-Souffrin 1990). For simplicity, we further fixed
$\sigma=1200$ km/s which is a typical value for double-peaked
emitters (see e.g. Strateva et al. 2003).

To physically constrain the fits, we assume that the Narrow Line
Region (NLR) is associated with the central
  component (r-system in Boroson \& Lauer 2009),
  and all narrow lines are Gaussian with the same Full Width at
  Half Maximum (FWHM). We also assume that both H$\alpha$ and
  H$\beta$ disk lines are from the same disk, so they must have the
  same inclination angle. We fitted the H$\alpha$ and
H$\beta$ lines from Chornock et al. (2009) in the following steps:
\begin{enumerate}
  \item We first fit a disk line to H$\alpha$, since it is less blended than
  H$\beta$. We subtract
  the disk line from the spectrum, and fit the
  residuals with 5 Gaussian profiles: one for the blue peak, two for the [N
II]$\lambda$6548 and 6583 (constrained to 1:3 height ratio), and two
(NLR, BLR) for the central peak. We then subtract the 5 Gaussian
profiles, and the residuals are considered to be the disk line
component, and we re-do the disk line fit to derive the inclination
for the disk component.
  \item We use the inclination angle derived in step 1, and fit a disk line
  to the H$\beta$ profile. We subtract the disk line from the spectrum,
  and fit the residual with 5 Gaussian profiles: one for the blue peak,
  two for [O III]$\lambda$ 4959 and 5007 lines respectively, and two
  (NLR, BLR) for the central peak. The central narrow H$\beta$ component and the two [O III]$\lambda$ 4959 and 5007
  lines are required to have the same FWHM.
\item We then use the FWHM derived in Step 2 to re-fit the
H$\alpha$ profile and require the central narrow H$\alpha$
component, [N II]$\lambda$6548 and 6583 (constrained to 1:3 height
ratio) to have the same FWHM as the [O III]$\lambda$ 4959 and 5007
lines (Figure 1).
\item Again, we used the inclination angle derived in step 3, and re-fit the H$\beta$ profile (Figure 2).
\end{enumerate}

\begin{figure}
\epsfig{file=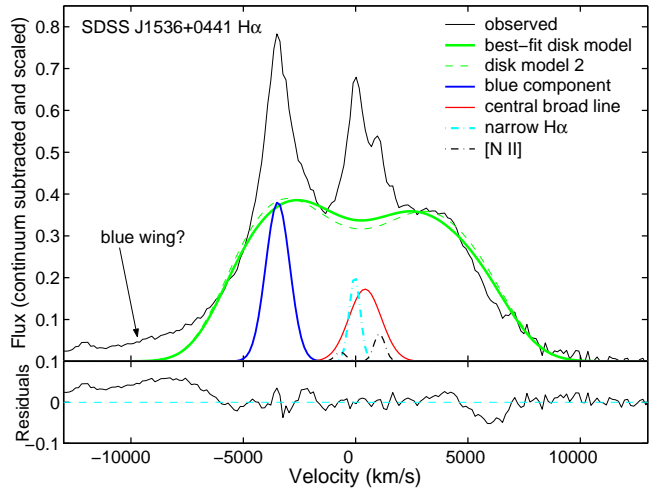, angle=0, width=\linewidth}
\caption{H$\alpha$ emission line profile of SDSS J1536+0441. In the
upper main panel, the black line is the continuum-removed spectrum
from Chornock et al. 2009, the solid green line is the best-fit disk
model line, the left blue line is the Gaussian fit to the blue peak,
the central red line is the Gaussian fit to the central peak
(r-system) with the narrow line component removed, and the cyan
dash-dotted line is Gaussian fit to the narrow $H{\alpha}$, and
black dash-dotted lines are Gaussian fits to [N II]$\lambda$6548,
6583 lines, with fixed FWHM given by fitting [O III]$\lambda$4959,
5007. The dashed green line shows the disk model line with outer
radius $8000\ r_G$, and the same inner radius and inclination as the
best-fit solid green line. The black solid line in the lower panel
shows the residuals.}
\end{figure}

\begin{figure}
\epsfig{file=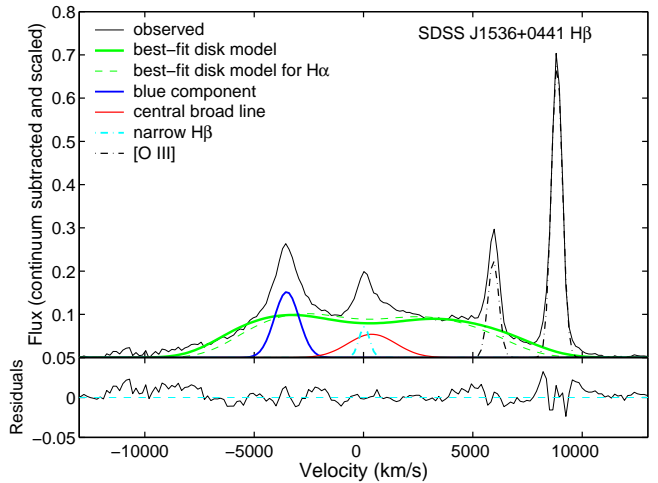, angle=0, width=\linewidth}
\caption{H$\beta$ emission line profile of SDSS J1536+0441. In the
upper main panel, the black line is the continuum removed spectrum
from Chornock et al. 2009, the solid green line shows the best-fit
disk model line with fixed inclination given in the fitting of
$H{\alpha}$ profile, the dashed green line shows the best-fit disk
model line for $H{\alpha}$ in Figure 1, the left blue line is the
Gaussian fit to the blue peak, the central red line is the Gaussian
fit to the central peak with the narrow line component removed, the
dash-dotted cyan lines are Gaussian fits to the narrow $H{\beta}$,
and the black dash-dotted lines are Gaussian fits to [O
III]$\lambda$4959, 5007 lines. The black solid line in the lower
panel shows the residuals.}
\end{figure}

\begin{table}
\begin{center}
\caption{Emission-line fit properties of SDSS
J1536+0441.\label{tbl-1}}
\begin{tabular}{lllll}
\tableline\tableline Components & Line (\AA) & Vel.\tablenotemark{a}
&
FWHM\tablenotemark{b} & Line Ratio\tablenotemark{c} \\
\tableline
Double-peaked & H$\alpha$ & N/A\tablenotemark{d} & 13580 & 8.5 \\
Double-peaked & H$\beta$ & N/A\tablenotemark{d} & 14890 & 5.7 \\
\tableline
Blue-system & H$\alpha$ & -3470 & 1340 & 1.0 \\
Blue-system & H$\beta$ & -3510 & 1330 & 1.0 \\
\tableline
Central broad line & H$\alpha$ & 420 & 1690 & 0.61 \\
Central broad line & H$\beta$ & 380 & 2570 & 0.69 \\
\tableline
Narrow lines & H$\alpha$ & -\tablenotemark{e} & 573\tablenotemark{f} & 0.24 \\
Narrow lines & [N II] 6548 & -\tablenotemark{e} & 573\tablenotemark{f} & 0.027\tablenotemark{g} \\
Narrow lines & [N II] 6583 & -\tablenotemark{e} & 573\tablenotemark{f} & 0.08\tablenotemark{g} \\
Narrow lines & H$\beta$ & -\tablenotemark{e} & 573\tablenotemark{f} & 0.20 \\
Narrow lines & [O III] 4959 & -\tablenotemark{e} & 573\tablenotemark{f} & 0.66 \\
Narrow lines & [O III] 5007 & -\tablenotemark{e} &
573\tablenotemark{f} & 2.0  \\
\tableline
\end{tabular}
\tablenotetext{a}{Velocities in km/s, negative value means
blueshifted and positive value means redshifted.}
\tablenotetext{b}{FWHM of the lines in km/s using the reduced
rest-frame spectra in Chornock et al. 2009.} \tablenotetext{c}{\
Line ratios are calculated compared with the blue component, i.e.
all the lines in Figure 1 (H$\alpha$) are compared with the line
flux of the blue H$\alpha$ component, and all the lines in Figure 2
(H$\beta$) are compared with the line flux of the blue H$\beta$
component.} \tablenotetext{d}{Not available since the centers of the
disk model lines are fixed in the fitting.}
 \tablenotetext{e}{The narrow lines are constrained within 5 \AA \ of
 their air wavelengths in the fitting, and it turned out all of them are consistent with zero
 velocity within 95\% confidence level.}
 \tablenotetext{f}{FWHMs of all the narrow
lines are constrained to be the same value in the fitting.}
 \tablenotetext{g}{[N II] 6548 and 6583 are constrained to 1:3 height ratio.}
\end{center}
\end{table}

The fitting results are shown in Figure 1 for H$\alpha$, Figure 2
for H$\beta$, and Table 1 for the best-fit parameters (with least
$\chi^2$). From the best-fit model, the disk inclination is
$i\sim47^o$. For the H$\alpha$ profile, the inner radius
$r_1\sim1000\ r_G$, and the outer radius $r_2\sim13000\ r_G$. For
the H$\beta$ profile, $r_1\sim800\ r_G$, and $r_2\sim7000\ r_G$. In
principle, the inner disk radius and the inclination determine the
width of the line, the inclination could be decoupled from the inner
radius from the net Gravitational redshift of the line, and the
outer radius determines the separation of the two shoulder-like
peaks (Chen \& Halpern 1989; Strateva et al. 2003). However, in the
case of SDSS J1536+0441, the shoulder-like peaks are obscured by the
blue peaks in both H$\alpha$ and H$\beta$, and blended by [O
III]$\lambda$4959 in the H$\beta$ profile, plus, there is a blue
wing component which cannot be fit well by an axisymmetric Keplerian
disk model, as shown obviously in the H$\alpha$ profile and also
present in H$\beta$ but weaker. As a result, the three remaining
parameters in the disk model, i.e. the inclination angle and the
inner and outer radii, are still degenerate to some degree, as shown
in Figure 3, and the errors are hard to estimate. Therefore, we did
not attempt to derive the errors in a statistical way, or try to get
the absolute value of the $\chi^2$ values of the fits. From the
H$\alpha$ profile fits, we cannot really distinguish model lines
with inner radius range from $800\ r_G$ with smaller inclination
angle ($i\sim40^o$) vs $1400\ r_G$ with larger inclination angle
($i\sim60^o$), as shown in Figure 3. We also cannot really
distinguish model lines with smaller outer radius like $8000\ r_G$
from model lines with larger outer radius like $13000\ r_G$, as
shown in the dashed line (disk model 2) and solid green line
(best-fit model line) in Figure 1. Therefore, the best-fit
parameters are only good as a rough initial estimate.

\begin{figure}
\epsfig{file=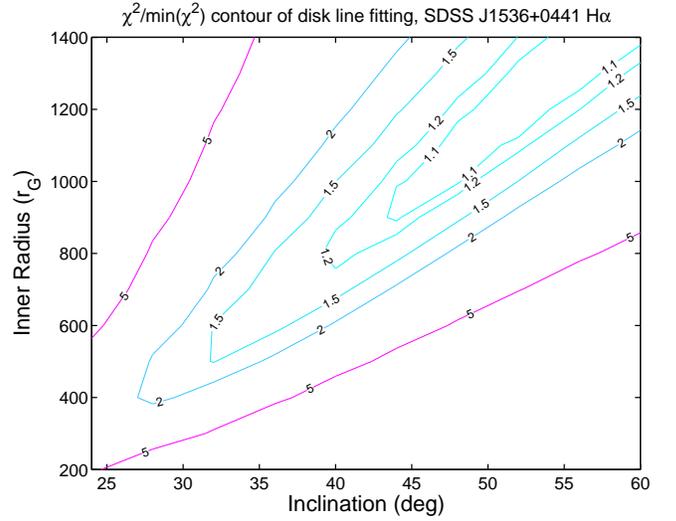, angle=0, width=\linewidth}
\caption{Re-scaled-$\chi^2$ contour of fittng H$\alpha$ emission
line profile of SDSS J1536+0441 in an axisymmetric Keplerian disk
model from Chen \& Halpern (1989). We only computed models with
inclination smaller than 60 degree, as is usually the case for type
I AGN.}
\end{figure}

Nevertheless, there are several very interesting features evident
from the circular disk model fitting:
\begin{enumerate}
\item As shown in Figures 1-2, both the H$\alpha$ and H$\beta$ profiles can be fit reasonably
well by a model with 4 components: a simple circular disk line from
the accretion disk around a central object (r-system), narrow lines
from a NLR associated with the central object (r-system), a central
broad line from the BLR associated with the central object
(r-system), and a broad blue line in another velocity field
(b-system). Only the first three components in the r-system are
normal in double-peaked emitters.
\item The H$\beta$ disk line is obviously wider than H$\alpha$, which
means H$\beta$ is emitted primarily from smaller radii in the disk.
As shown in Figure 2, the best-fit disk line for the H$\beta$ is
shown in solid green, while the best-fit disk line for H$\alpha$ is
shown in dashed green. Here FWHM(H$\alpha$)=0.91FWHM(H$\beta$),
which is consistent with the higher excitation temperature for
H$\beta$, and is consistent with the value found in normal
(non-double-peaked emitter) AGNs (Stirpe 1990; Kaspi et al. 2000).
\item The central broad component is redshifted compared with the
narrow lines by $\sim400$ km/s, indicating there might be some form
of inflow. Fe II emission and intermediate-width H$\beta$ components
are found in SDSS AGNs to be systematically red-shifted by about 400
km/s with respect to the narrow lines, which may come from an
intermediate-line region located at the outer portion of the BLR
which is dominated by inflow (Hu et al. 2008a, 2008b). The existence
of such an intermediate-width broad line region might be common in
AGNs (see e.g. Brotherton et al. 1994; Zhu et al. 2009). Another
alternative explanation is gravitational redshift. Eracleous \&
Halpern (1994) found in a radio-loud AGN sample that the broad
H$\alpha$ lines at half-maximum are preferentially redshifted by
$<\Delta \lambda/\lambda>=(6\pm2)\times10^{-4}$ on average, i.e.
$\sim 180$ km/s. If the 400 km/s redshift in the central broad
component is caused only by gravitational plus transverse redshift,
then it corresponds to Keplerian motion at a radius of $\sim 10^3\
r_G$ and a Keplerian velocity of $\sim 10^4$ km/s, which is much
larger than the FWHM of the central broad lines. Therefore, it is
more likely come from some form of inflow.
\item There is a blue wing component, strong in the H$\alpha$
profile and weaker in the H$\beta$ profile, with velocity as high as
$\sim10,000$ km/s and indicating some form of outflow. Such a
component has been seen in many double-peaked emitters (e.g. 3
double-peaked AGNs out of a sample of 8 in Eracleous \& Halpern
2003, showed a blue wing excess), and could be the wind from the
accretion disk (Elvis 2000).
\end{enumerate}

\section{Binary black hole as a model to explain SDSS J1536+0441}
As discussed in Lauer \& Boroson (2009), the sharp-peaked blue
component in SDSS J1536+0441 is different from what is seen in known
double-peaked emitters, which are usually more flattened and
shoulder-like. This blue component, which was the key characteristic
stimulating the study of the system and was discussed extensively in
previous papers (Boroson \& Lauer 2009 and references therein),
could possibly be fit by an extremely asymmetric disk model, where
the blue peak comes from a high emissivity region (such as a hot
spot) moving towards us. However, how to produce and maintain such a
huge hot spot is not clear.

Instead of invoking a highly-asymmetric disk model, alternatively,
we propose that SDSS J1536+0441 is a binary black hole system, where
the disk-line component comes from the accretion disk around the
primary black hole, the central component comes from the more
spherical BLR around the primary SMBH, and the blue component comes
from the region around the secondary SMBH, which is moving towards
us right now. Our model is illustrated in Figure 4.

\begin{figure}
\epsfig{file=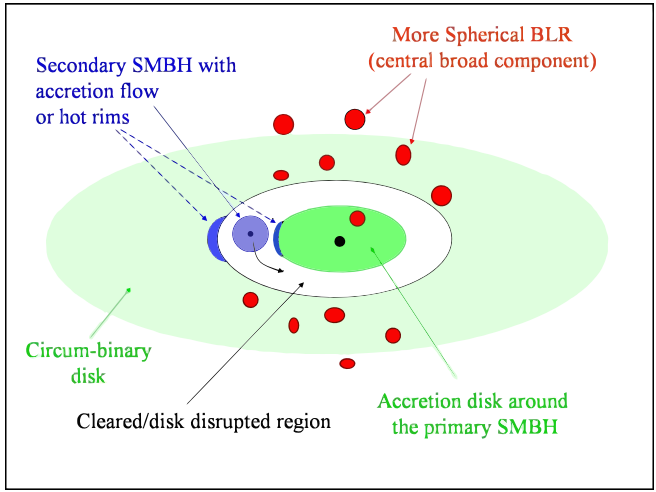, angle=0, width=\linewidth}
\caption{Illustration of our model for SDSS J1536+0441, where the
secondary SMBH spirals in and clears a gap in the gas disk around
the primary SMBH (not drawn to scale). The secondary is currently
moving towards us. In the context of the model proposed here, the
disk-line component comes from the accretion disk around the primary
black hole, the central component comes from the more spherical BLR
around the primary black hole, and the blue component comes from
either the accretion flow around the secondary SMBH, or the rims of
the circum-primary and the circum-binary disks which are close to
and heated by the tidal perturbations of the secondary.}
\end{figure}

The blue peaks could come from the accretion flow onto the secondary
SMBH. The possible existence of two broad line systems as the
consequence of two merging black holes has been predicted by Gould
\& Rix (2000), one from the primary and one from the accretion disk
around the secondary. Fe II, which is believed to be photoionized
and probably from winds flowing off the surfaces of accretion disks
of AGNs (Baldwin et al. 2004; Vestergaard \& Peterson 2005), has
been found primarily associated with the blue-system velocity
(Boroson \& Lauer 2009; Lauer \& Boroson 2009), and supports the
existence of an accretion disk around the secondary.

The blue peaks could also come from the rims of the circum-primary
and the circum-binary disks, which are close to and heated by the
tidal perturbations of the secondary SMBH. Lodato et al. (2009)
showed in their simulations that the tidal torques from the
secondary SMBH heat the edges of the gap, creating bright rims
around the secondary, which could account for as much as 50\% of the
integrated light from the disk. SDSS J1536+0441 would be a perfect
example if this is the case. Note the velocity separation between
the blue peak and the central peak ($v \sim 3500$ km/s) is similar
to the Keplerian velocity of the disk outer radii. If we assume the
inclination angle as $i=47^o$, the separation velocity is the
projected orbital velocity of the secondary, and the secondary is
just moving towards us, then its orbital radius is $2(c\sin i/v)^2
\sim 8000 \ r_G$, which is consistent with the outer radius in our
H$\alpha$ (see the disk model 2 (dashed green line) in Figure 1) and
H$\beta$ profile fits.

Though there is one double-peaked emitter, 3C 390.3, in the
reverberation mapping AGN sample, which follows the Luminosity-BLR
size relation (hereafter L-R relation) well (Peterson et al. 2004;
Kaspi et al. 2005), it is under debate whether the empirical L-R
relation derived from reverberation mapping results could be applied
to double-peaked AGNs (Wu \& Liu 2004; Zhang et al. 2007). Zhang et
al. (2007) found that the black hole masses of double-peaked
emitters estimated from pure stellar velocity dispersion (Lewis \&
Eracleous 2006) are about six times smaller than the virial black
hole masses using the reverberation mapping L-R relation and FWHMs
of H$\beta$ lines, though in some AGNs the two masses are consistent
while in some others the virial mass could be 10-20 times larger. If
we assume a typical starlight fraction value of 0.33 for
double-peaked emitters (Eracleous \& Halpern 2003), and all the
other continuum is from the primary SMBH, then use the FWHM of the
best fit double-peaked H$\beta$ line, we could derive a virial black
hole mass following the reverberation mapping L-R relation (Kaspi et
al. 2005). We adopt a flat cosmology with $H_0=71$ km/s/Mpc and
$\Omega_m=0.27$), and find the size of the primary BLR (which is the
accretion disk here) $R=141$ lt-days, $M=4.6\times 10^9 \ M_\odot$.
If we adopt the value of six as the factor of the mass increase by
the virial calculation, then the mass of the primary is about
$M_1=10^9 \ M_\odot$. It is difficult to estimate the mass of the
secondary. As discussed in section 2, the central broad component is
redshifted by $\sim400$ km/s. If it is due to the orbital motion of
the primary SMBH rather than inflow, while the blue broad component
is due to the orbital motion of the secondary, then the mass ratio
will be $\sim0.1$ and the secondary SMBH mass will be $M_2=10^8 \
M_\odot$. However, this is unlikely to be the case, since the disk
line component agrees well with the redshifts of the narrow lines,
and therefore the secondary to primary mass ratio must be smaller.
Alternatively, if we assume the blue broad component is from
virialized gas around and illuminated only by the secondary, and the
continuum luminosity of the secondary is one-tenth the value of
total continuum (which is roughly the line flux ratio between the
blue Gaussian component and the double-peaked disk line in
H$\alpha$), then following the L-R relation, the virial mass of the
secondary black hole is about $M_2=10^7 \ M_\odot$. Note that the
black hole masses we estimated are similar to the values derived by
Boroson \& Lauer (2009) ($M_1=10^{8.9} \ M_\odot$, $M_2=10^7.3 \
M_\odot$) in a somewhat different way.

\section{Discussion}
Since we wish to make as few assumptions as possible, and since both
binary SMBHs and double-peaked emitters are rarely observed, why do
we include both? We answer this question as follows.

1. SDSS J1536+0441 is likely a double-peaked emitter. The red
shoulders in the emission lines shown in Chornock et al. (2009), and
the fact that it is fit well by our simple disk model line by
excluding both the blue and central peaks, suggest there are
emissions from the disk.

2. There is no simple model that works well for SDSS J1536+0441. The
sharp blue peak has not been seen before in any double-peaked
emitter. Lauer \& Boroson (2009) listed three that may be analogous
to SDSS J1536+0441, but none of them seem convincing. Therefore, if
we accept the hypothesis that SDSS J1536+0441 is a double-peaked
emitter, it must be a peculiar one, which then requires additional
unusual assumptions.

3. The existence of a minor black hole is physically more natural
than an extremely asymmetric accretion disk. In fact, we do expect
SMBH binaries as natural consequences of galaxy mergers (Begelman et
al. 1980; Civano et al. 2009; Comerford et al. 2009).

 4. Double-peaked emission lines are not common and most ($\sim 97\%$ at
$z<0.33$) AGNs in the SDSS sample do not show double-peaked line
profiles (Strateva et al. 2003). Instead of asking why a few percent
of AGNs show double-peaked emission lines, we ask why most AGNs do
not show double-peaked emission lines, since virtually all AGNs must
have accretion disks. Possible reasons include: i) a vertical
extended ionizing structure is required for double-peaked emitters
(Chen \& Halpern 1989), so if there is no illuminating source
shining on the disk, the double-peaked lines will either disappear
or be weak and buried on the continuum, or ii) if the disk extends
to large radius (Rokaki et al. 1992; Jackson et al. 1991), iii) the
disk is face-on (Corbin 1997) or iv) there is a disk wind (Murray \&
Chiang 1997; Eracleous et al. 2004). Any of the above could smear
out the double-peaked profile. We note that the existence of a minor
black hole helps produce disk emission lines in two aspects. First,
when the secondary spirals in, the outer edge of the inner disk will
form a density spike and puff up (Armitage \& Natarajan 2000).
Therefore the covering factor of the disk becomes larger; and a hot
vertical extended torus (Chen \& Halpern 1989) or outflow might also
be formed due to the disturbance brought by the minor black hole
(Armitage \& Natarajan 2000), which helps to illuminate the disk. As
discussed in Eracelous \& Halpern (2003), such a vertical extended
structure will also help explain the preferential association of
double-peaked emitters with radio-loud AGNs (as shown in Strateva et
al. 2003 that double-peaked AGNs are 1.6 times more likely to be
radio sources), since the original vertical extended ion torus model
(Rees et al. 1982) was intended to explain the formation of radio
jets. Second, when the secondary black hole approaches merger with
the primary black hole, the disk is truncated at a radius of the
order of the binary separation, therefore the double-peaked profile
is not smeared.

Since SDSS J1536+0441 can not be unique as the only double-peaked
emitter in a binary supermassive black hole system, such systems
might also exist in many known double-peaked emitters, where the
tidal torques from the secondary black hole clear the outer region
of the disk around the primary black hole, and might also help
illuminate the disk. This leads to the question: why do other
double-peaked emitters not show any evidence of the existence of a
secondary black hole?

There are two possible reasons. First, when the secondary is much
smaller than the primary, its BLR might be totally stripped, and its
heating on the edges of the disks is not strong enough to produce
excess emission line components. Therefore, the emission from the
accretion disk around the primary dominates. In such cases, the
secondary might excite spiral wave or other instabilities in the
accretion disk around the primary and cause the variability of the
double-peaked emission lines, such as seen in 3C 390.3 (Chakrabarti
\& Wiita 1993; Eracleous et al. 1997; Gilbert et al. 1999).
Moreover, about half of the double-peaked emitters require some form
of ``non-axisymmetries'', and how to produce the asymmetries is
challenging since a significant perturbation is needed. The
secondary will naturally bring asymmetries to the system. In such
cases, we would also expect these double-peaked emitter AGNs to be
more variable than the average, which is the case for optical X-ray
variability on timescales of years (see e.g. Strateva et al. 2008).
Second, the reason why we haven't identified any double-peaked
emitter binary system yet is in order to find the signature of a
secondary black hole, we need long-term observations, which are
generally not available (but could be from the Digital Access to a
Sky Century @ Harvard (DASCH) project - see Grindlay et al. 2009).
Several double-peaked emitters have been monitored for two decades
(Halpern \& Filippenko 1988; Eracleous et al. 1997; Gezari et al.
2007 and references therein), and in most of them, there is no
evidence for the systematic drift signal from a binary system.
However, in one system, Arp 102B, which is the prototype disk
emitter, there is an excess component which oscillated between the
red and blue peaks of the disk line from 1991 to 1994, and from 1999
to 2003 (Newman et al. 1997; Gezari et al. 2007). During both
epochs, the period is about 2 yrs, and the velocity drift amplitude
is about 8000 km/s from 1991 to 1994, and about 6000 km/s from 1999
to 2003. Therefore, if both oscillations are caused by an orbiting
bright spot, the two derived masses of the central SMBH are
discrepant by a factor of 2 (Gezari et al. 2007). We suggest that it
could be a SMBH binary, and the periodic drift excess components are
from the heated rims caused by the orbital motion of the secondary,
while the contributions from the inner and outer rims are different
in the two epochs. The transient nature of the excess could be due
to the instabilities in the system (Lodato et al. 2009). The flux
ratio from the two rims likely also changes over a timescale of
several orbits. Therefore the derived amplitude of the velocity of
the excess component also changes. Here the oscillation period could
be the orbital period of the secondary, the velocity drift from 1991
to 1994 could be dominated by the orbital motion of the heated inner
rim near the secondary, and the velocity drift from 1999 to 2003
could be dominated by the orbital motion of the heated outer rim
near the secondary which has relatively smaller amplitude.

Another question is, are all binary black hole AGN systems
double-peaked emitters? Our answer is they are not. The stage that
both black holes have BLRs is likely to be short lived. The inner
disk around the primary might be depleted soon since the secondary
acts like a dam and the matter in the outer circum-binary disk can
hardly get through the gap swept by the secondary (Lodato et al.
2009; but see Artymowicz \& Lubow 1996 for gas streams penetrating
the disk gap and supply mass to the binary). A similar SMBH binary
model has been proposed by Bogdanovic et al. (2009) to explain
another multiple-redshifts system, SDSS J092712.65+294344.0
(hereafter SDSS J0927). SDSS J0927 was discovered by Komossa et al.
(2008) as a candidate for a recoiling black hole, and is the other
system found in the search by Boroson \& Lauer (2009), which shows
two sets of narrow lines and only the blue one has associated broad
lines. Bogdanovic et al. (2009) proposed that the system is a SMBH
binary surrounded by a circum-binary disk (Ivanov et al. 1999;
Cuadra et al. 2009), where the primary is inactive and has no BLR,
while the secondary moving towards us is accreting and emits broad
lines from its accretion disk and narrow lines from the inner rim of
the circum-binary disk, which might resemble the later evolutionary
stage of SDSS J1536+0441 in our model.

If we adopt $M_1=10^9\ M_\odot$, $M_2=10^7\ M_\odot$, and a
Keplerian projected velocity $3500$ km/s of the secondary at
inclination $47^o$, then the separation of the two black holes is
$0.38$ pc and the period is 490 yr. The decay timescale for the
binary to merge due to gravitational radiation alone is
$1\times10^{12}$ yr (Peters 1964). However, the gas disk around the
black holes will help solve the so called final parsec problem, but
the timescale given in hydrodynamic simulations is quite uncertain
and depends on the mass ratio between the disk and the secondary
(Lodato et al. 2009; Cuadra et al. 2009). If we assume that all
double-peaked emitters are merging binary systems in the gas
dissipation stages, then there will be $\geq3\%$ nearby AGNs in
merging binary systems. This means, on average, these AGNs spend
$\geq 4\times10^8$ years in merging. If there is one merger event
per lifetime, then each merger takes $\geq 4\times10^8$ years in its
gas dissipation phase, which is roughly a few times the disk's
viscous timescale, and so consistent with the hydrodynamic
simulation results when the disk mass is larger than or comparable
with the secondary mass (Armitage \& Natarajan 2002; Lodato et al.
2009).

For our estimated 490 years for the orbital period of SDSS
J1536+0441, we may not be able to see the velocity change of the
blue peak over several years. Spectroscopic monitoring on timescales
of decades could provide the most direct test of the binary
hypothesis. A possible test of our model in the near future might be
UV observations from refurbished \emph{HST}. As discussed in
Chornock et al. (2009), the Ly $\alpha$ and C IV lines in
double-peaked emitters frequently lack double-peaked velocity
profiles (Halpern et al. 1996; Eracleous et al. 2004). Eracleous et
al. (2004) suggested that the double-peaked low-ionization Balmer
lines are from the dense accretion disk where the Ly $\alpha$
emission is suppressed, and the single-peaked Ly $\alpha$ line is
from a lower density higher-ionization wind. So under the picture of
Eracleous et al. (2004), there are three possible cases:
\begin{enumerate}
\item If the blue peaks in the H$\alpha$ and
H$\beta$ lines in SDSS J1536+0441 are from the accretion disk of a
single SMBH (as in Chornock et al. 2009), then a blue peak is not
likely to be present in the Ly $\alpha$ profile;
\item If the blue
peaks in the H$\alpha$ and H$\beta$ lines in SDSS J1536+0441 are
from region around the secondary SMBH, i.e. either accretion flow
around the secondary or rims of the inner circum-primary and outer
circum-binary disks, but there is no lower density,
higher-ionization wind from the secondary, then a blue peak is not
likely to be present in the Ly $\alpha$ profile;
\item If the blue
peaks in the H$\alpha$ and H$\beta$ lines in SDSS J1536+0441 are
from region around the secondary SMBH, and there is a lower density,
higher-ionization wind from the secondary (possibly arisen from the
accretion disk around the secondary), then a blue peak will be
present in the Ly $\alpha$ profile.
 \end{enumerate}
Therefore, the presence of blue peaks shifted by $\sim 3500$ km/s in
the Ly $\alpha$ and C IV emission lines would strongly support our
model (case 3); the absence of multiple velocity peaks in the Ly
$\alpha$ and C IV emission lines, however, would still leave the
question open (neither case 1 nor case 2 could be ruled out).

\acknowledgments We are grateful to the anonymous referee for
helpful comments and suggestions. S.T. thanks Martin Elvis, Shuang
Nan Zhang, Yuexing Li for stimulating discussions and Lei Hao for
valuable suggestions. We thank Ryan Chornock for sending us the
spectra data in their paper.

\end{document}